\documentclass[aps,twocolumn,showpacs]{revtex4}
\usepackage[dvips]{graphicx}
\usepackage{amsmath}
\usepackage{amsfonts}
\usepackage{amssymb}
\usepackage[linktocpage,breaklinks=true,colorlinks=true,linkcolor=black,filecolor=black,citecolor=black,pagecolor=black,urlcolor=black]{hyperref}

\usepackage{color}

\newcommand{\eq}[1]{\begin{equation}{#1}\end{equation}}

\newcommand{\kB}{k_\mathrm{B}}

\begin{document}

\title{Light induced crystallization of cold atoms in a thin 1D optical tube}
\date{\today}

\author{Tobias Grie{\ss}er}
\affiliation{Institut f\"ur Theoretische Physik, Universit\"at Innsbruck, Technikerstra{\ss}e~25, 6020~Innsbruck, Austria}

\author{Helmut Ritsch}
\email{Helmut.Ritsch@uibk.ac.at}
\affiliation{Institut f\"ur Theoretische Physik, Universit\"at Innsbruck, Technikerstra{\ss}e~25, 6020~Innsbruck, Austria}

\begin{abstract}

Collective off resonant scattering of coherent light by a cold gas induces long-range interactions via interference of light scattered by different particles. In a 1D configuration these interactions grow particularly strong for particles trapped  along an optical nanofiber. We show that there exists a threshold pump laser intensity, above which the gas can be found in a crystalline, selfsustained order. In the nonabsorbing regime we determine the critical condition for the onset of order as well as the forms of particle density and scattered field patterns along the fiber above threshold. Surprisingly, there can coexist multiple stationary solutions with distinct density and field profiles. 
\end{abstract}

\pacs{37.30.+i, 37.10.-x, 51.10.+y}

\maketitle
The astonishing experimental accomplishments in the optical control and manipulation of cold atomic gases in the past decade now allow to deterministically load them in extremely well controlled optical traps of almost any shape down to effectively zero temperature. A particular fruitful example are periodic optical lattices in which many intriguing phenomena of solid state physics can be studied with unprecedented control\cite{bloch2008many}.

In contrast to conventional solids, however, the spatial order and lattice geometry is fixed by the external lasers and does not appear from a self consistent dynamics of particle interactions. As the back-action of the atoms onto the confining light fields is generally negligible\cite{asboth2008optomechanical}, local lattice perturbations do not propagate and long range interactions, which appear as phonons in solids are absent. This changes for spatially tightly confined fields as in small optical resonators or optical micro-structures\cite{domokos2002quantum}. 

Experimentally it is very challenging to implement and load microtraps close to optical microstructures, where such atom field coupling is enhanced\cite{colombe2007strong, alton2010strong, schleier2011optomechanical}. In an important step Rauschenbeutel and coworkers, however, recently managed to trap atoms in an array of optical dipole traps generated by two color evanescent light fields alongside a tapered optical fiber\cite{vetsch2010optical} where the backaction of even a single atom on the propagating fiber field is surprisingly strong\cite{domokos2002quantum,chang2012cavity}. This setup was improved with higher control and coupling by other groups recently\cite{goban2012demonstration}. With the atoms firmly trapped within the evanescent field of the fiber modes, field mediated atom-atom interaction and collective coupling to the light modes play a decisive role in this a setup\cite{zoubi2010hybrid}.

Already a decade ago it was theoretically predicted\cite{domokos2002collective} and experimentally confirmed\cite{black2003observation,baumann2010dicke,arnold2012self} that light scattering within optical resonators induces self-ordering of atoms in regular patterns maximizing collective coupling to the cavity mode\cite{ritsch2012cold,gopalakrishnan2009emergent}. This transition can be directly monitored from the super-radiant light scattering \cite{black2003observation,arnold2012self} and appears also at zero temperature in the quantum regime as phase transition from a superfluid to a supersolid\cite{baumann2010dicke}. In an recent proposal Chang and coworkers predicted, that nanofiber-mediated infinite range dipole-dipole coupling can also induce stable regular patterns of laser illuminated atoms trapped along the fiber\cite{chang2012self}. The stable configurations, characterized by minimal dipole interaction energy, assume surprising configurations and exhibit characteristic collective light scattering. 

\begin{figure}[h!]

\includegraphics[width=8cm]{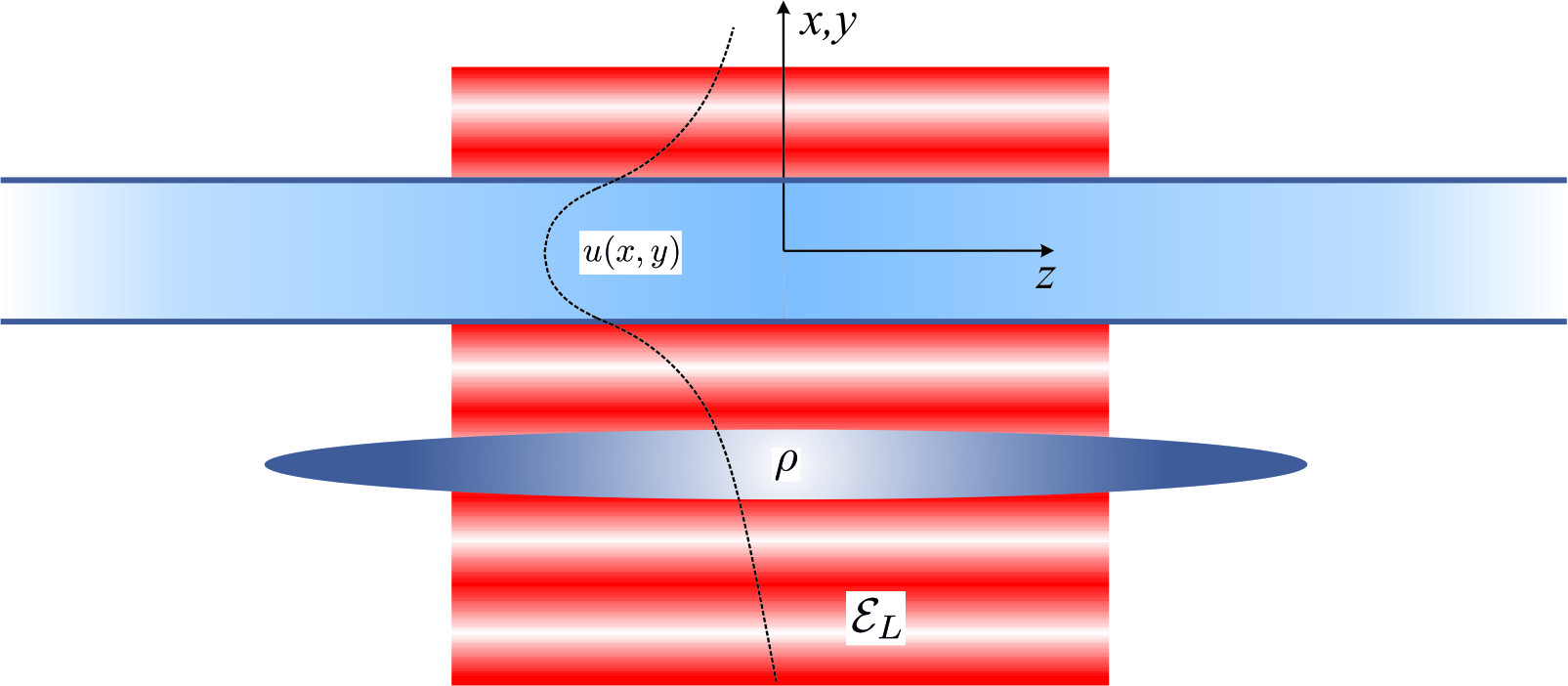}
\caption{Cigar-shaped atomic gas alongside optical nanoguide}\label{fig1}
\end{figure}

Here we develop a generalized many particle model to study the properties of such a light scattering induced crystallization of an laser illuminated ultracold gas in an elongated 1D trap at finite temperature as schematically drawn in Fig.\ref{fig1}. The atoms trapped parallel to the fiber are illuminated at right angle by a Gaussian laser beam of frequency $\omega_L$, sufficiently detuned from any atomic resonance so that spontaneous emission plays only a minor role and their polarizability $\alpha$ has only a negligible imaginary part.

The laser field $\mathbf E_L=(\mathcal E_L(\mathbf x)e^{-i\omega_Lt}+c.c.)\mathbf e_L$ gives rise to the field $\mathbf E_s$ scattered by the atoms. Neglecting polarization effects and retardation, it can be written as $\mathbf E_s=(\mathcal E_s(\mathbf x,t)e^{-i\omega_Lt}+c.c.)\mathbf e_L$, its envelope satisfying Helmholtz's equation
\begin{subequations}\label{VlasovHelmholtz}
\eq{\label{FullHelmholtz}\nabla^2\mathcal{E}_s +k_L^2\big(n_F^2+\chi\big)\mathcal{E}_s=-k_L^2\chi\mathcal{E}_L.}
Here, $n_F(\mathbf x)$ denotes the refractive index profile of the optical fiber and $\chi(\mathbf x,t) ={\alpha}\rho(\mathbf x,t)/{\epsilon_0}$ is the susceptibility of the particles. The particle density distribution $\rho$ is obtained from the momentum integral 
$\rho(\mathbf x,t)=\int F(\mathbf x,\mathbf p,t)d^3p$ 
over the one-body distribution function $F(\mathbf x,\mathbf p,t)$ proper to the gas. 
In the classical mean-field limit it satisfies Vlasov's equation
\eq{\label{FullVlasov}
\frac{\partial F}{\partial t}+\frac{\mathbf p}{m}\!\cdot\!\frac{\partial F}{\partial \mathbf  x}-\frac{\partial }{\partial \mathbf x}\left(\Phi_d+U_T\right)\!\cdot\!\frac{\partial F}{\partial \mathbf  p} =0.}
\end{subequations}
Here, $U_T$ denotes the elongated dipole trap potential and \eq{\Phi_d=-\alpha \left|\mathcal{E}_s+\mathcal E_L\right|^2} the optical potential due to pump laser and fiber field.  For strong radial confinement of the gas the one-body distribution approximately factorizes into a longitudinal and a transverse part,
$F(\mathbf x,\mathbf p,t)\simeq f(z,p_z,t) F_\perp(\mathbf x_\perp,\mathbf p_\perp)$, where $F_\perp$ is the Maxwell-Boltzmann distribution for the transverse degrees of freedom: $F_\perp:= Z_\perp^{-1}\exp\left[-\beta\left(\frac{\mathbf p_\perp^2}{2m}+U_\perp(\mathbf x_\perp)\right)\right]$. $\beta$ denotes the inverse thermal energy and $U_\perp$ the radial confining potential.

In the absence of atoms the fiber supports only a single relevant TE mode, which -- within the framework of scalar theory -- is supposed to possess a propagation constant $\beta_m$ and a normalized transverse mode function $u(x,y)$ extending outside the fiber \cite{vetsch2010optical,chang2012self}. The dominant forces on the particles along $z$  are due to photon scattering into and out of the fiber. As long as the total atomic susceptibility stays small even for large particle numbers, the radial fiber mode function is only weakly perturbed and we can set $\mathcal E_s(\mathbf x,t)\simeq\sqrt{A}\,E(z,t)u(x,y)$ with the cross section $A:=\left(\int d^2x_\perp d^2p_\perp u^2 F_\perp \right)^{-1}$. Integrating over the transverse degrees of freedom we finally arrive at an effective description of the longitudinal dynamics
\begin{subequations}\label{HV}
\eq{\label{Helmholtz}\frac{\partial^2E}{\partial z^2}+\left(\beta_m^2+k_L^2\tilde\chi\right)E=-k_L^2 \tilde\chi E_L,}
\eq{\label{Vlasov}\frac{\partial f}{\partial t}+\frac{p_z}{m}\frac{\partial f}{\partial z}-\frac{\partial }{\partial z}\!\left(U-\alpha\!\left[|E|^2+2E_LE_r\right]\right)\!\frac{\partial f}{\partial p_z} =0,}
\end{subequations}
where $E_r$ is the real part of $E$. The effective laser field is given by $E_L:=\sqrt{A}\,\int d^2x_\perp d^2p_\perp\mathcal{E}_L u F_\perp$ and the local atom-fiber coupling is governed by the effective susceptibility
\eq{\label{Susc}\tilde\chi(z,t):=\frac{\alpha}{\epsilon_0 A}\int^{\infty}_{-\infty} f(z,p_z,t)dp_z}
proportional to the atomic line density. To ensure physically consistent solutions, equation \eqref{Helmholtz} for the electric field has to fulfill Sommerfeld's radiation conditions
\begin{equation}
\label{BC}\frac{\partial E}{\partial z}=\pm i\beta_m E(z,t),~~z\rightarrow \pm\infty,
\end{equation}
which ensure outgoing waves at infinity. 

Equations \eqref{HV} can at least numerically be solved directly. At this point we nevertheless restrict ourselves to such stationary solutions, which correspond to a gas in thermal equilibrium\cite{asboth2005self}, where the particles are distributed according to the Maxwell-Boltzmann distribution
\eq{\label{ThermEq}f(z,p_z)=Z^{-1} e^{-\beta(p_z^2/2m+U)}\,e^{\beta\alpha\left(|E|^2+2E_pE_r\right)}.} 

Substituting the effective susceptibility \eqref{Susc} obtained from a thermal distribution \eqref{ThermEq} into the effective Helmholtz equation \eqref{Helmholtz} leads to a highly nonlinear equation, which determines the selfconsistent electric field. We assume here that the external part of the potential, $U(z)$, can be harmonically approximated by $U(z)=\frac{1}{2}m\omega_z^2 z^2$ within the (very large) thermal extension of the gas $l_z:=2\kB T/m\omega_z^2\gg\beta_m^{-1}$, and we will ignore the dependence of the driving laser amplitude on position, i.e. $E_L(z)=E_L$.
 
With these assumptions one sees that the zero field solution $E(z)\approx 0$ and
\eq{\label{ThermEq0}f(z,p_z)\equiv f_0(z,p_z)=Z^{-1} e^{-\beta(p_z^2/2m+U)}}
always solves Helmholtz's equation, which we will call the normal phase.

\begin{figure}
	\centering
		 \includegraphics[width=8cm]{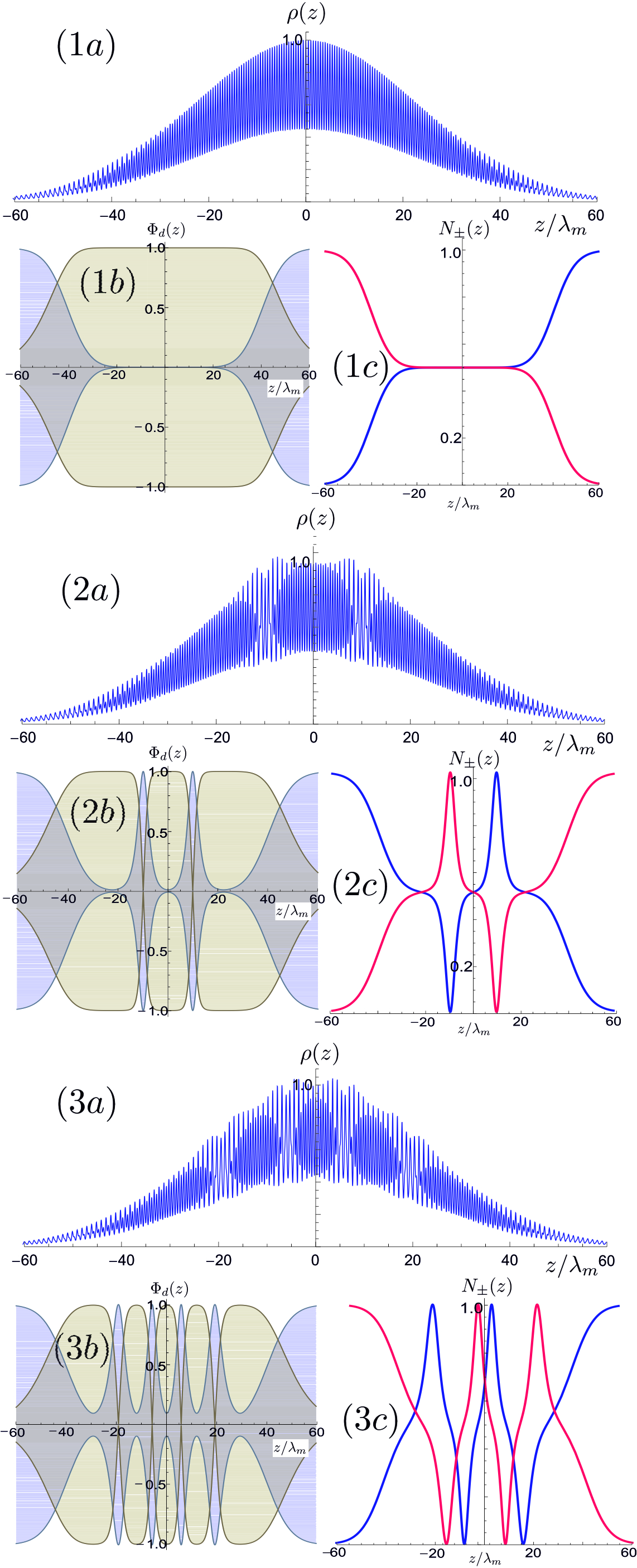}
	\caption{Properties of three coexisting selfordered solutions on different branches $n=0$ ($(1a)-(1c)$), $n=1$ ($(2a)-(2c)$) and $n=2$ ($(3a)-(3c)$) in the strong collective coupling regime for $\zeta_0=150/\pi$ and $\varepsilon=9.5\varepsilon_c$.  The upper plots (a) show the atomic density along the fiber, while the plots labeled by (c) depict the local fraction of right $N_+$ (blue curve) and left  $N_-$  (red curve) traveling photons, the number of zeros being equal to the branch number $n$. The plots labeled (b) depict the envelopes of the two parts of the optical potential. Blue indicates the potential from scattering between pump laser and fiber, the brown area shows the part due to photon redistribution within the fiber. For $n=0$ the latter dominates in the center of the cloud causing a density modulation with a period of one half of $\lambda_m=2\pi/\beta_m$. This region is broken up into three for $n=1$ and into five, for $n=2$.}\label{Bild2}
\end{figure}
Before we investigate the possibility of selfordered solutions, let us examine the dynamical stability of the normal phase, using the Helmholtz-Vlasov equations \eqref{HV}, linearized around the unordered state
\begin{subequations}\label{linVlas}
\eq{\frac{\partial^2E}{\partial z^2}+\beta_m^2E=-E_L\frac{k_L^2\alpha}{\epsilon_0A}\int^{\infty}_{-\infty}f_1(z,p_z,t)dp_z ,}
\eq{\frac{\partial f_1}{\partial t}+\frac{p_z}{m}\frac{\partial f_1}{\partial z}-\frac{d U}{d z}\frac{\partial f_1}{\partial p_z}=-2\alpha \frac{\partial E_L E_r}{\partial z}\frac{\partial f_0}{\partial p_z}.}
\end{subequations}
Here, $f_1(z,p_z,t)$ represents a small deviation from the normal phase, such that the complete distribution function is given by $f(z,p_z,t)=f_0(z,p_z)+f_1(z,p_z,t)$.
We seek solutions to \eqref{linVlas}, which are of the form
\begin{subequations}\label{NormalMode}
\eq{f_1(z,p_z,t)=\psi(z,p_z)e^{st}+\psi^*(z,p_z)e^{s^*t},}
\eq{E(z,t)=a(z)e^{st}+b^*(z)e^{s^*t}.}
\end{subequations}
In this Ansatz, $(\psi,a,b)$ will be referred to as normal mode and $s=\gamma+i\omega$ as the complex valued mode parameter.
Whenever $\gamma>0$, the deviation from equilibrium defined by the normal mode increases exponentially in time, causing the eventual destruction of the normal phase. The existence of such a mode implies dynamical instability. With the help of asymptotic expansions, we find that if \eqref{NormalMode} is to solve \eqref{linVlas} and satisfy the radiation conditions, the mode parameter must satisfy $D_n(s)=0$ for some $n\in\mathbb{Z}$, where
\eq{D_n(s)=(2n+1)\pi+i\frac{k_L^2\alpha^2}{\epsilon_0A}\iint^{\infty}_{-\infty}\frac{E_L^2\,\partial f_0/\partial p_z}{s+i\beta_m v_z} dp_z dz.}

This condition first demands $\omega=0$. Introducing the collective coupling parameter $\zeta_0$ and effective pump strength $\varepsilon$
\eq{\zeta_0:=\frac{k_p}{\beta_m}\frac{N\alpha}{A\lambda_L\epsilon_0}, \; \varepsilon:=\frac{\alpha E_L^2}{k_BT},}
we then see that there exists at least one normal mode with a positive growth rate $\gamma>0$, if and only if
\eq{\label{CC1}\varepsilon>\frac{1}{2\zeta_0}=:\varepsilon_c.}

Hence as our first important result we see that beyond this pump threshold  \eqref{CC1} the normal phase ceases to be stable and no longer represents a physical state of the system. We therefore study other solutions of \eqref{HV} for $\varepsilon>\varepsilon_c$.
Fortunately, we can obtain approximate solutions with the help of perturbation theory in the weak collective coupling regime, $\zeta_0\leq1$, as well as in the strong collective coupling regime $\zeta_0\gg1$. The real and imaginary parts of the electric field envelope $E_r, E_i$ \eq{E_{r,i}(z)=a_{r,i}(z)\cos(\beta_m z+\phi_{r,i}(z))E_L,} can be assumed to behave almost harmonically with phases $\phi_r, \phi_i$ and amplitudes $a_r, a_i$ that vary slowly on a scale defined by $\beta_m^{-1}$.  
This amplitudes and phases are directly related to the complex amplitudes of the more familiar decomposition into left and right running waves, $E=(E_+e^{i\beta_mz}+E_-e^{-i\beta_mz})E_L$ via $E_\pm=a_r e^{\pm i\phi_r}+ia_i e^{\pm i\phi_i}$.
A perturbative analysis of the steady state Helmholtz equation reveals, that 
$\Theta:=\frac{a_r^2+a_i^2}{2}=\frac{1}{2}(|E_+|^2+|E_-|^2)$, proportional to the sum of locally right going and left going photons, is spatially constant and will thus serve us as order parameter.

Let us first consider the weak collective coupling regime, where scattering from the laser into the fiber and vice versa dominates over scattering within the fiber.  The demand to satisfy the radiation conditions \eqref{BC} leads to the equation determining this order parameter

\eq{\label{OP}2\zeta_0 I_1\left(2\varepsilon\sqrt{\Theta}\right)=(1+2n)\sqrt{\Theta} I_0\left(2\varepsilon\sqrt{\Theta}\right),~~n\in\mathbb{N}_0,}
where $I_k$ denotes the $k$th modified Bessel function of the first kind.

\begin{figure}
	\centering
		\includegraphics[width=8cm]{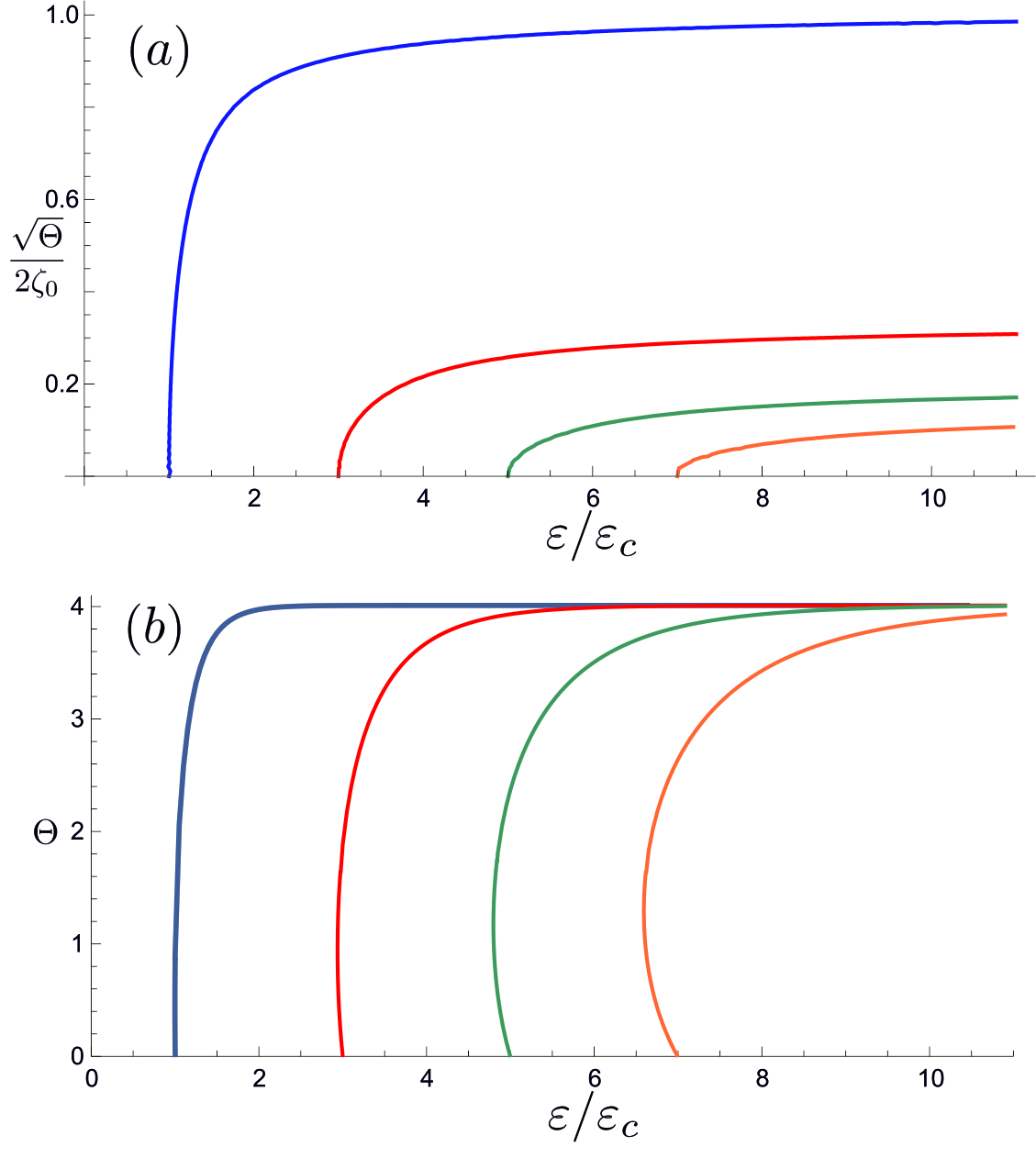}

	\caption{Branches of the order parameter vs energy ratio according to \eqref{OP} and \eqref{OP1} for $n=0,\ldots,3$ in the weak collective coupling regime $(a)$, with $\zeta_0=0.05$ and in the strong collective coupling regime $(b)$, with $\zeta_0=75/\pi$. Note that in the latter regime all branches converge and there may be even more than $n$ families of solutions for $\varepsilon_c<\varepsilon<(1+2n)\varepsilon_c$.}\label{Bild3}
\end{figure}

As quite a surprise one observes that the solution is not unique and for
$\varepsilon_c<\varepsilon<\frac{1+2n}{2\zeta_0}$,
there exist $n$ different solutions ${\Theta_n}$ as shown in the example of figure \ref{Bild3}. As the phase of the outgoing scattered light is not determined, each solution of \eqref{OP} corresponds to an infinite family of solutions of \eqref{HV} with the periodic density modulation slightly displaced under a slow envelope. The appearance of this degeneracy of course corresponds to breaking of a continuous symmetry.  It is important to note that the point of first appearance of an ordered family of solutions, $\varepsilon=\varepsilon_c$, exactly coincides with the point where the normal phase becomes unstable. All this is confirmed by a numerical solution of the underlying equations \eqref{HV}.
Focusing on the region close to the first branching point, we find the behavior of the order parameter to be given by \eq{\Theta\sim\varepsilon-\varepsilon_c,~~\zeta_0<1}
and
\eq{\Theta\sim\left(\varepsilon-\varepsilon_c\right)^{1/2},~~\zeta_0=1.}
Curiously, if $\zeta_0>1$, the order parameter is discontinuous, exhibiting a jump of magnitude
\eq{\Delta\Theta=16(\zeta_0-1).}
The results of the perturbation theory hold only as long as $\Theta\ll1$ and the predicted jump soon violates this assumption.
Let us therefore turn to the strong collective coupling regime. Again, each possible value of the order parameter corresponds to an infinite family of stationary solutions and
any nonzero value for $\Theta$ satisfies
\eq{\label{OP1}\zeta_0 \varepsilon\pi=4(1+2n)P(\Theta)K[P^2(\Theta)\Theta^2],~~n\in\mathbb{N}_0,}
where $P(\Theta):=\frac{1+2\varepsilon \Theta}{4+6\varepsilon \Theta}$ and $K$ denotes the complete elliptic integral of the first kind.
In this regime, there are at least $n$ families of ordered solutions whenever $1<2\zeta_0\varepsilon<1+2n$. For an example see figure \ref{Bild3}.
Furthermore, we find that according to perturbation theory, the order parameter is bounded by $\Theta<4$. Figure \ref{Bild2} depicts some properties of the solutions corresponding to possible values of the order parameter. It should be noted that while we have here shown the existence of multiple families of selfordered thermal solutions of the Vlasov-Helmholtz equations, nothing has been said about their dynamical stability properties, which remains an open problem.
\par
Certain phenomena called optical binding, similar to the ordering discussed in this work, have been observed with laser illuminated small beads in liquids\cite{burns1990optical} in 1D and even 2D geometries as well. More interestingly, in analogy with cavity induced selfordering we expect a corresponding phase transition also in the zero temperature limit, where the threshold is determined by the recoil energy and effective particle interaction strength replacing thermal energy. This should have distinctly different properties as compared to a 1D prescribed optical lattice. Obviously, even without the presence of the fiber, the interference of the scattered field by two distant atoms can induced long range forces and ordering. As part of the scattered field is lost, this leads to a higher threshold, but to some extend the particles itself can guide the scattered light. In the limit of sufficient initial particle density the trapped atoms themselves could form a fiber like guide for the scattered light enhancing long range interactions. One could speculate that the trapping field for the 1D confinement could be unnecessary and be provided by the light guided by the atoms. The predicted ordering phenomena could also be related to recent observations of collective scattering in dense atomic vapors\cite{greenberg2011bunching,schmittberger2012free} where light induced selfordering mechanisms play an important role.
\par
We thank A. Rauschenbeutel, J. Kimble and I.Cirac for stimulating discussions and acknowledge support by the Austrian Science Fund FWF through the projects SFB FoQuS P13.

\bibliographystyle{apsrev}
\bibliography{Selforganization_with_Fiber_2013}\section*{Appendix}
In this appendix we indicate, how the results of the present work can be obtained in a systematic manner by means of canonical perturbation theory, not restricting ourselves to thermal equilibrium. Helmholtz's equation with a stationary gas density $\rho$ in the source term reads
\eq{\label{Helmholtz1}\frac{\partial^2E}{\partial z^2}+\beta_m^2E=-\frac{k_L^2\alpha}{\epsilon_0 A}(E+ E_L) \rho.}
It is important to note that for \emph{any} stationary solution of Vlasov's equation \eqref{Vlasov}, which necessarily takes the form $f(z,p_z)=F(p^2/2m+\Phi_d+U)$, where $\Phi_d=-\alpha(|E|^2+2E_LE_r)$, we have for the corresponding density $\rho(z)=g[\Phi_d+U]$ for some real function $g$. Let $G$ denote the antiderivative of $g$, i.e. $G'=g$, and let us define the canonical momenta $\Pi_{r,i}:=\partial_z E_{r,i}$. Then, the above equations for the real and imaginary parts of the selfconsistent field are the canonical equations of the following Hamiltonian
\eq{H=\frac{1}{2}\left[\Pi_r^2+\Pi_i^2+\beta_m^2(E_r^2+E_i^2)-\frac{k_L^2}{\epsilon_0 A}G[\Phi_d+U]\right],}
which has the physical interpretation of being proportional to the energy density of the scattered electric field.
It can be written as the Hamiltonian of two uncoupled resonant oscillators, $H_0$, plus a coupling, $H=H_0+H_1$, where
\eq{H_1=-\frac{k_L^2}{2\epsilon_0 A}G[\Phi_d+U],}
which part we shall consider a perturbation. Let us also introduce action-angle variables $(J_r,J_i, \psi_r,\psi_i)$ for the unperturbed part via the prescription
\begin{subequations}
\eq{E_{r,i}=\sqrt{\frac{2 J_{r,i}}{\beta_m}}\sin(\psi_{r,i})}
\eq{\Pi_{r,i}=\sqrt{2\beta_m J_{r,i}}\cos(\psi_{r,i}).}
\end{subequations}
Then we find the transformed Hamiltonian as $H=H_0(J)+\epsilon H_1(J,\psi,\epsilon z)$, where now $H_0(J)=\beta_m(J_r+J_i)$ and we have introduced a purely formal expansion parameter $\epsilon$. The explicit dependence on $z$ is due only to the external trap $U(z)$ and thus weak. Next we will seek a canonical transformation, effected by the mixed variables generating function $S=S(\bar J,\psi,z)$ to new variables $(\bar J, \bar\psi)\equiv (\bar J_r, \bar J_i, \bar \psi_r, \bar \psi_i)$ according to
\eq{J=\nabla_\psi S,~~\bar \psi=\nabla_{\bar J}S}
and leading to the new Hamiltonian $K=H+\frac{\partial S}{\partial z}$, which we wish to make as simple as possible.
To exploit the assumed smallness of $H_1$, we express the generating function as
\eq{S=\bar J \cdot \psi+\epsilon S_1+\epsilon^2 S_2+\ldots.}
Likewise expanding $K=K_0+\epsilon K_1+\ldots$, we find
\begin{subequations}
\eq{K_0(\bar J)=H_0(\bar J),}
\eq{\label{Eq2}K_1(\bar J, \bar \psi, \epsilon z)=\beta_m\sum_{k=r,i}\frac{\partial S_1}{\partial \bar\psi_k}+H_1(\bar J, \bar \psi,\epsilon z),}
\end{subequations}
and so forth. Setting $\epsilon=1$ and introducing the unperturbed trajectories $\bar\psi_{r,i}(\zeta):=\bar\psi_{r,i}+\zeta$, we find
\begin{multline}
S_1(\bar J,\bar\psi(\zeta),z)-S_1(\bar J,\bar\psi,z)=\\=
\frac{1}{\beta_m}\int_0^\zeta \left[K_1(\bar J,\bar\psi(\zeta'),z)-H_1(\bar J,\bar\psi(\zeta'),z)\right]d\zeta'.
\end{multline}
Imposing the periodicity of the generating function in the angles leads to the solvability condition
\eq{\label{Solv}\int_0^{2\pi} \left[K_1(\bar J,\bar\psi(\zeta),z)-H_1(\bar J,\bar\psi(\zeta),z)\right]d\zeta=0,}
which must be satisfied by any potential $K_1$.
Now we use our freedom to choose $K_1$ to be a function of the phase difference only, $K_1=\tilde K_1(\bar J, \bar \Delta,z)$,  where $\bar\Delta:=\bar\psi_r-\bar\psi_i$. Then it is easy to see, that \eqref{Solv} demands
\eq{\tilde K_1(\bar J, \bar \Delta,z):=\frac{1}{2\pi}\int_0^{2\pi} H_1(\bar J, \zeta+\bar\Delta,\zeta,z)d\zeta.} In this way we end up with the new and simplified Hamiltonian, to first order in the perturbation given by
$K(\bar J, \bar \psi,z)=H_0(\bar J)+\tilde K_1(\bar J, \bar \Delta,z)$.
As this new Hamiltonian depends, by construction, only on the phase difference $\bar\Delta$, we immediately obtain an integral of motion, $\bar\Theta:=\bar J_r+\bar J_i$, proportional to the order parameter, which is the total number of quanta of excitation in the left and right propagating field components. Therefore, the number of degrees of freedom is reduced to just two, namely $\bar\Delta$ and $\bar D:=\bar J_r-\bar J_i $, whose equations of motion are the canonical equations pertaining to the Hamiltonian
\eq{\bar H(\bar D,\bar\Delta,z):=2\tilde K_1\left(\frac{\Theta+\bar D}{2},\frac{\Theta-\bar D}{2},\bar \Delta,z\right)}
and can be solved in the weak and strong collective coupling limits respectively.
\end{document}